\documentclass[pra,amsmath,amssymb,notitlepage, twocolumn,superscriptaddress,longbibliography,nofootinbib,floatfix]{revtex4-1}
\usepackage[pdftex,colorlinks=true,linkcolor=darkblue,citecolor=blue,urlcolor=darkred]{hyperref}
\usepackage{braket}
\usepackage{amsmath,amsfonts, amssymb, amsthm, dsfont}
\usepackage{yfonts}
\usepackage{bm,breqn}
\usepackage{mathrsfs}
\usepackage{array,makecell}
\usepackage{svg}
\usepackage{graphicx}
\usepackage[caption=false]{subfig}
\usepackage{verbatim,soul}

\usepackage{tikz}
\usetikzlibrary{calc}
\usepackage{multirow}
\usepackage{color}
\usepackage[capitalise]{cleveref}

\theoremstyle{definition}

\newtheorem*{lemma*}{Lemma}

\definecolor{darkblue}{rgb}{0.,0.,0.4}
\definecolor{darkred}{rgb}{0.5,0.,0.}

\graphicspath{{figs/}{}}

\begin{document}
\title{Black Mirrors: CPT-Symmetric Alternatives to Black Holes}
\author{Kostas Tzanavaris}\affiliation{\Edinburgh}
\author{Latham Boyle}\affiliation{\Edinburgh}\affiliation{\PI}
\author{Neil Turok}\affiliation{\Edinburgh}\affiliation{\PI}
\newcommand*{\Edinburgh}{Higgs Centre for Theoretical Physics, James Clerk Maxwell Building, Edinburgh EH9 3FD, UK} 
\newcommand*{\PI}{Perimeter Institute for Theoretical Physics, Waterloo, Ontario N2L 2Y5, Canada}

\begin{abstract}
Einstein's equations imply that a gravitationally collapsed object forms an event horizon.
But what lies on the other side of this horizon?  In this paper, we question the reality of the conventional solution (the black hole), and point out another, topologically distinct solution: the \textit{black mirror}.  In the black hole solution, the horizon connects the exterior metric to an interior metric which contains a curvature singularity.  In the black mirror, the horizon instead connects the exterior metric to its own CPT mirror image, yielding a solution with 
smooth, bounded curvature.  We give the general stationary (charged, rotating) black mirror solution explicitly, and also describe the general black mirror formed by gravitational collapse.  The black mirror is the relevant stationary point when the quantum path integral is equipped with suitably CPT-symmetric boundary conditions, that we propose.  It appears to avoid many vexing puzzles which plague the conventional black hole.
\end{abstract}
\maketitle

\section{Introduction}

The Einstein-Rosen bridge (the first wormhole geometry) was put forth by Einstein and Rosen in 
\cite{Einstein:1935tc}. 
One of their motivations was to get rid of the apparent ``singularity" at the event horizon $r=2m$ by introducing a new coordinate $u$ by $r=2m+u^2$, and letting $u$ extend over $-\infty<u<\infty$. This
describes two Schwarzschild exteriors (regions $I$ and $I'$ in Fig.~\ref{fig-kruskal-1}), glued at 
the radius $r=2m$ (the midpoint in Fig.~\ref{fig-kruskal-1}). However, it was later deemed to be unphysical as the papers of Oppenheimer-Snyder \cite{Oppenheimer:1939ue} describing spherical collapse showed that the singularity at $r=2m$ is just a coordinate singularity, and an infalling observer following the collapsing matter will not feel anything unusual happening at the event horizon.  This (and other developments) led to the modern belief that the black hole interior really exists, and that the black hole's horizon is completely non-singular.

Nevertheless, the modern black hole picture of gravitational collapse is not without its problems. Aside from the emergence of curvature singularities, there are other problems that the currently accepted theories cannot address adequately, including the breakdown of causality, {\it i.e.}, the emergence and stability of Cauchy horizons \cite{Dafermos:2017dbw}, as well as the information \cite{Hawking:1976ra} and firewall \cite{Almheiri:2012rt} paradoxes. 

In this paper, we argue that certain natural modifications of Einstein-Rosen bridges -- which we call {\it black mirrors} -- 
are, in fact, direct consequences
of CPT symmetry, 
and offer novel solutions to the problems mentioned above, along with other merits discussed below.\footnote{Also note several other proposals for variations to the usual black hole geometry or topology \cite{Gibbons:1986rzu, Saravani:2012is, Visser:1989kh, Visser:1989kg, tHooft:2016qoo, tHooft:2016rrl, Carroll:2009maa, Loran:2010qn, Loran:2010zy, Koiran:2024nng, Bouhmadi-Lopez:2019kkt, Brahma:2021xjy} (and see Appendix \ref{CPT_appendix}).}

\section{The Schwarzschild Black Mirror}
\label{Schwarzschild}

For simplicity, let us first consider the Schwarzschild metric, describing an eternal, static, uncharged object.  In Schwarzschild coordinates $(t,r,\theta,\varphi)$ it is
\begin{equation}
  \label{Scwharzschild}
  ds^{2}=-f(r)dt^{2}+\frac{dr^{2}}{f(r)}+r^{2}d\Omega^{2}
\end{equation}
where $f(r)\equiv 1-(2m/r)$ and $d\Omega^{2}\equiv d\theta^2+{\rm sin}^{2}\theta\,d\varphi^{2}$.
This solution has a horizon at $r=2m$, the root of $f(r)$. 
Note that Schwarzschild coordinates only cover the exterior region $r>2m$, and become singular at the horizon $r=2m$ (which lies at $t=\pm\infty$ in these coordinates). 

\begin{figure}
\begin{center}
\includegraphics[height=0.8in,width=1.6in]{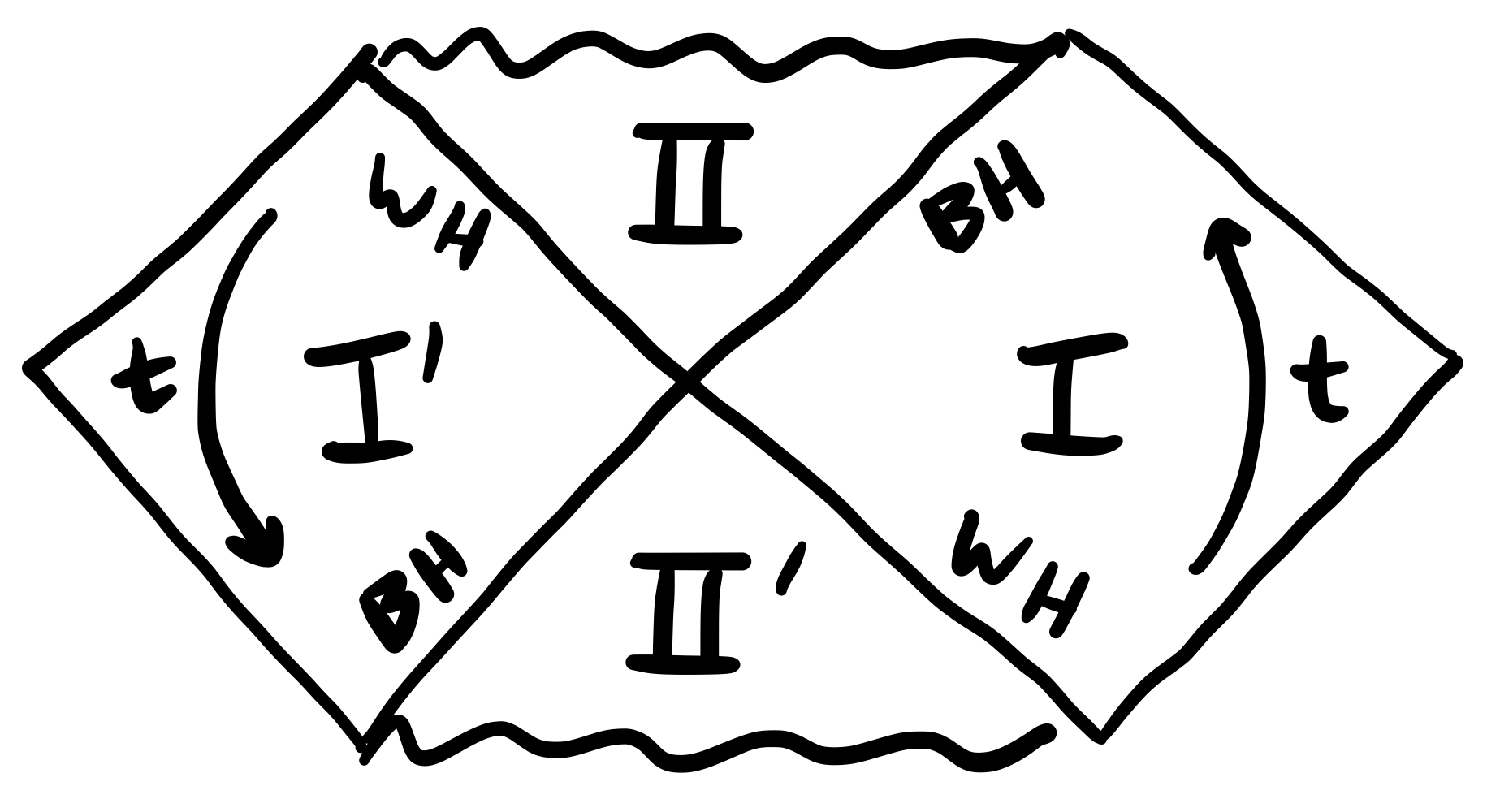}
\caption{Penrose diagram for the eternal Schwarzschild (Kruskal) black hole, indicating the two exteriors ($I$, $I'$), two interiors ($II$, $II'$), two black horizons (BH), two white horizons (WH), and the time-like killing vector $\partial_{t}$, which points upward in exterior $I$, and downward in exterior $I'$.} 
\label{fig-kruskal-1}
\end{center}
\end{figure}

The conventional view is that the {\it full} spacetime (including the horizon, and what lies behind it) is the maximal analytic extension obtained by switching from Schwarzschild to Kruskal coordinates \cite{dInverno}.  The Penrose diagram of this extension is shown in Fig.~\ref{fig-kruskal-1}: it has two exterior regions $I$, $I'$, two interior regions $II$, $II'$, two black horizons (BHs) that particles can only enter, and two white horizons (WHs) that particles can only exit.

\begin{figure}
\begin{center}
\includegraphics[height=0.9in,width=2.4in]{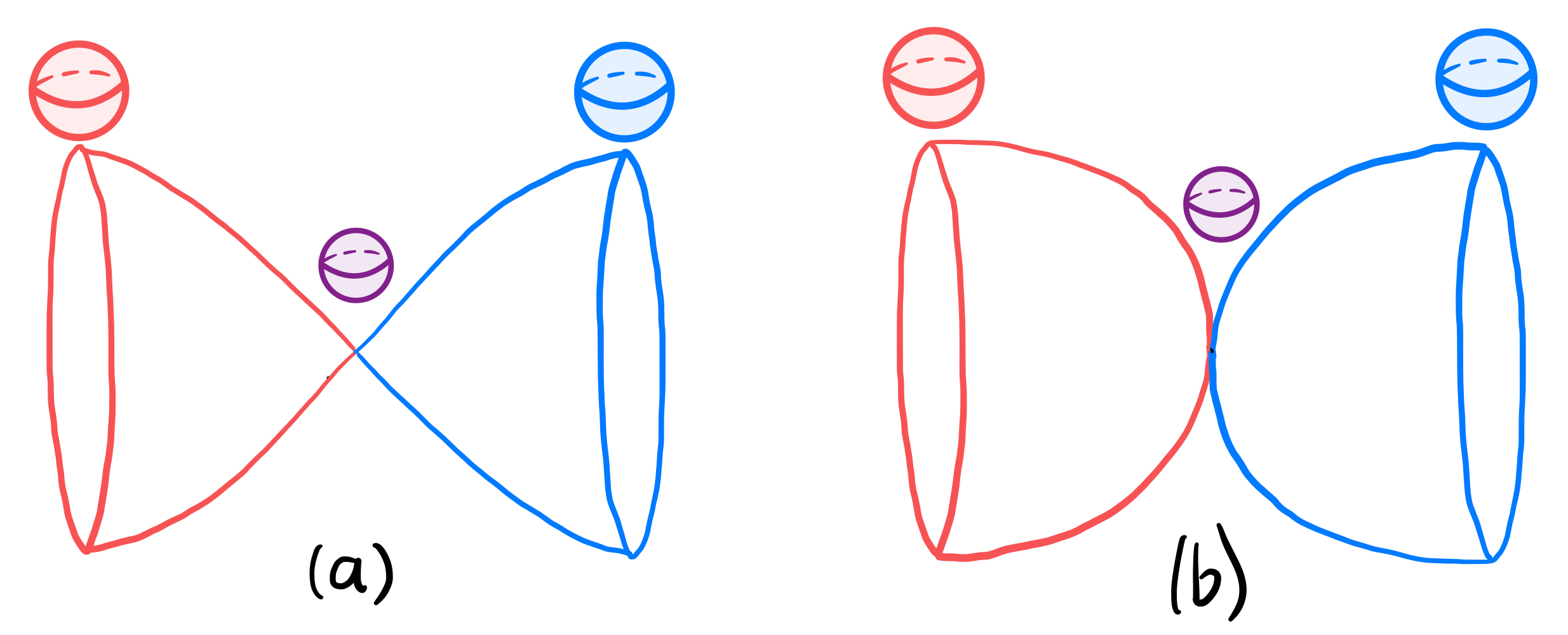}
\caption{The Euclidean Schwarzschild metric, with $\sigma>0$ in blue, and $\sigma<0$ in red.  (a) for generic $\tau$ period: conic singularity; (b) for special $\tau$ period $8\pi M$: no conic singularity.}
\label{euclidean-bh}
\end{center}
\end{figure}

Following the classic Gibbons-Hawking analysis \cite{Gibbons:1976ue}, in order to extract the quantum/thermodynamic properties of this spacetime, and to define quantum field theory on it, it is standard to start in Schwarzschild coordinates (the coordinates that make manifest the full isometry group of the metric) and then ``Wick rotate" to the \textit{Euclidean} Schwarzschild metric, by replacing the Schwarzschild coordinates, $t$ and $r$, by the Euclidean time $\tau$ and Euclidean radius $\sigma>0$, defined by
\begin{equation}
    \label{r(sigma)}
    t(\tau)=-i\tau,\qquad r(\sigma)=2m\Big[1+\big(\frac{\sigma}{4m}\big)^{2}\Big].
\end{equation}
So the Euclidean Schwarzschild metric (in $\tau,\sigma$ coordinates) is given by
\begin{equation}
    \label{EuclideanSchwarzschild}
    ds^{2}=\frac{2m}{r(\sigma)}\Big(\frac{\sigma}{4m}\Big)^{2}
    d\tau^{2}+\frac{r(\sigma)}{2m}d\sigma^{2}
    +r(\sigma)^{2}d\Omega^{2}.
\end{equation}
For small $\sigma$ (which corresponds to the near-horizon limit in Lorentzian signature), this becomes
\begin{equation}
    \label{EuclideanSchwarzschildApprox}
    ds^{2}\approx\frac{\sigma^{2}}{(4m)^{2}}
    d\tau^{2}+d\sigma^{2}+(2m)^{2} d\Omega^{2}.
\end{equation}
One recognizes this as the metric of a 2-dimensional ($\tau,\sigma$) cone $C^2$,  with a 2-dimensional ($\theta,\varphi$) sphere $S^2$ of radius $2m$ living over each point of the cone (see Fig.~\ref{euclidean-bh}a).  The $C^2$ has angular coordinate $\propto\tau$, and radial coordinate $\sigma$ measuring the proper distance from the tip $\sigma=0$.  Note that, in the standard Euclidean analysis \cite{Gibbons:1976ue}, one considers the {\it one-sided} $C^2$ (the blue cone in Fig.~\ref{euclidean-bh}a), corresponding to the black hole exterior region $I$ ($\sigma>0$).   

Then, if the Euclidean time coordinate's period is precisely $8\pi m$ ({\it i.e.}\ if we identify $\tau\sim\tau+8\pi m$), the conic singularity at $\sigma=0$ is removed, and the one-sided $C^2$ (the blue cone in Fig.~\ref{euclidean-bh}a) becomes a non-singular 2D Euclidean surface (the blue surface in Fig.~\ref{euclidean-bh}b).  From this, one infers that the black hole's temperature $T_{BH}$ is the inverse of this natural period: $T_{BH}=1/8\pi m$ \cite{Gibbons:1976ue}. 

After this elegant argument, we appear to be left with a smooth, complete Euclidean spacetime (the blue surface in Fig.~\ref{euclidean-bh}b) which is usually regarded as the entire Euclidean Schwarzschild metric.  But this blue surface corresponds purely to the Schwarzschild {\it exterior} $\sigma>0$ (region $I$ in Fig. \ref{fig-kruskal-1}a). What happened to the spacetime {\it behind} the event horizon at $\sigma=0$? 

To answer this question, let us first take a step back to the conical geometry of Fig.~\ref{euclidean-bh}a.  Here it is clear that the single (blue) cone naturally extends to the double (blue+red cone) by letting $\sigma$ extend to negative values, just as Einstein and Rosen did \cite{Einstein:1935tc} on the Lorentzian side, with their coordinate $u=\sigma/\sqrt{8m}$.  Then, when we identify $\tau\sim\tau+8\pi m$ (removing the conical singularity), we obtain the full (blue+red) Euclidean Schwarzschild metric, Fig.~\ref{euclidean-bh}b.  Note: this full geometry still has nothing corresponding to the black hole {\it interior}; instead, it has two {\it exteriors} glued together at $\sigma=0$ (an $S^2$ of area $16\pi m^2$, corresponding to the horizon).

Further evidence for the two-sided geometry in Fig.~\ref{euclidean-bh}b comes from black hole entropy.  Gibbons and Hawking \cite{Gibbons:1976ue} compute the entropy of the Schwarzschild black hole by computing the action of the Euclidean Schwarzschild metric (which comes entirely from the Gibbons-Hawking-York boundary term \cite{York:1972sj, Gibbons:1976ue}).  And, because they compute this action for the {\it one-sided} (blue) spacetime, they obtain a non-zero answer.  This is a clue that the one-sided Euclidean metric is physically incomplete. \textit{If, instead, we compute the action/entropy for the two-sided Euclidean spacetime, the answer vanishes (as the two boundary terms, on isometric but oppositely oriented boundaries, make equal and opposite contributions, see Appendix \ref{saddle-point-app}).}  This suggests a picture in which the full (blue+red) geometry corresponds to the full (pure, zero entropy) quantum state, while the half (blue) geometry corresponds to the mixed state obtained by tracing out the other (red) half; and the black hole entropy corresponds to the entanglement entropy ({\it i.e.}\ the von Neumann entropy of the blue side, after tracing out the red side).  This in turn explains why the entropy is proportional to the area of the horizon (since the entanglement entropy between two regions is characteristically proportional to the size of the boundary separating them).

\begin{figure}
\begin{center}
\includegraphics[height=3in,width=2in]{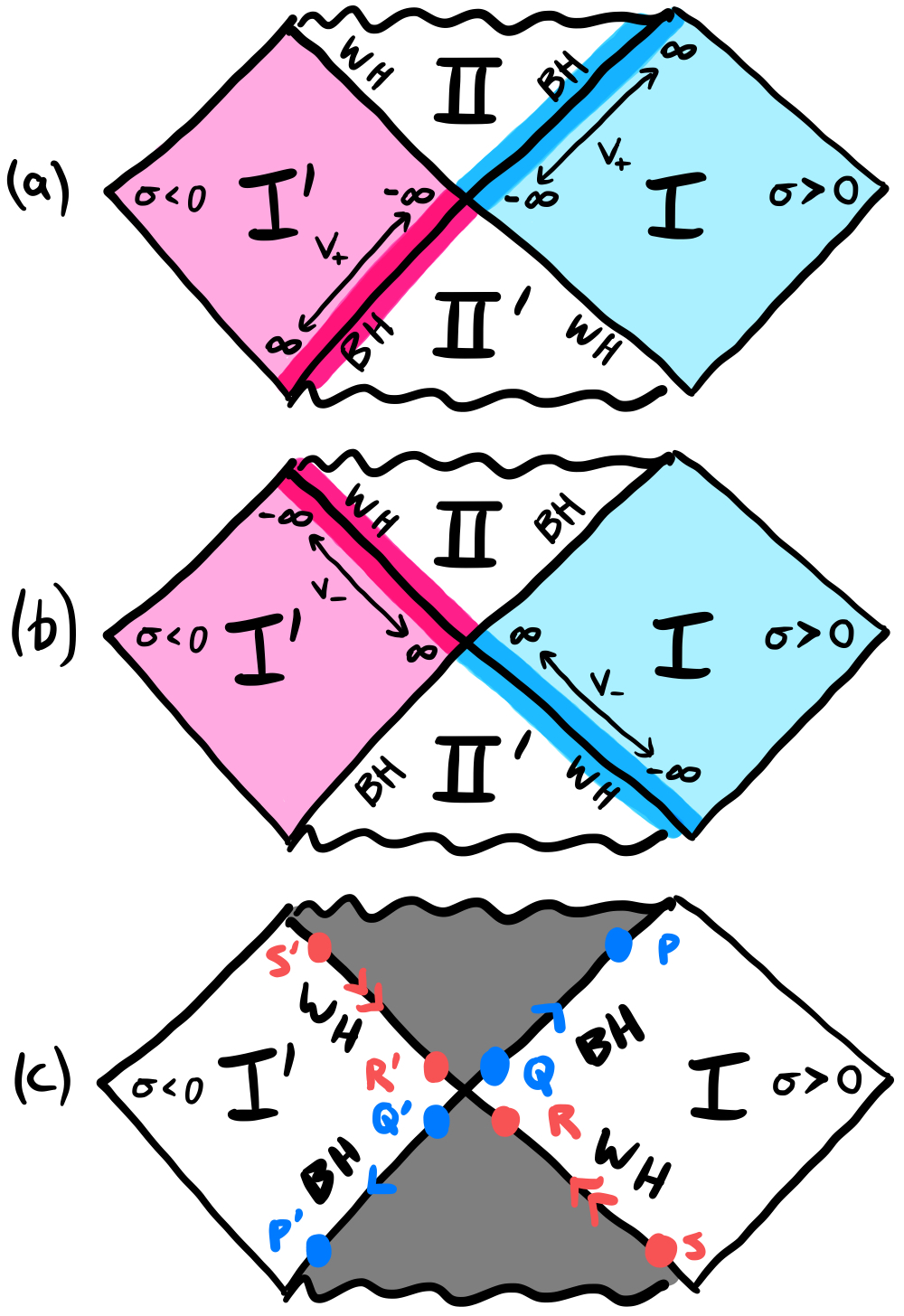}
\caption{(a) $(v_+,\sigma)$ coordinates; (b) $(v_-,\sigma)$ coordinates; (c) the Penrose diagram for the eternal black mirror: exteriors I, I' are glued together, BH to BH and WH to WH, as indicated ({\it e.g.} $P\sim P'$, $R\sim R'$), and with antipodal inversion of the $(\theta,\varphi)$ 2-sphere (see Appendix \ref{CPT_appendix}), so the interiors II, II' and their curvature singularities (dark grey regions) are removed.}
\label{stationary-mirror}
\end{center}
\end{figure}

If we Wick rotate the full Euclidean metric (\ref{EuclideanSchwarzschild}) back to Lorentzian signature using $\tau=it$, we obtain a metric that now covers the full exterior $r>2m$ in Fig.~\ref{fig-kruskal-1}a ({\it i.e.}\ regions $I$ and $I'$).  This is the Einstein-Rosen bridge \cite{Einstein:1935tc}, which is incomplete as it stands, and needs to be extended to a complete spacetime.  The conventional completion is the Kruskal extension of Fig.~\ref{fig-kruskal-1}; but here we suggest a different completion. 
The simplest way to introduce this alternative possibility is to extend the metric across the BH (WH) by replacing $t$ with the ingoing (outgoing) Eddington-Finkelstein time coordinate $v_+$ ($v_-$), related by $dv_{\pm}=dt\pm\frac{dr}{f(r)}$, so that the metric 
becomes
\begin{equation}
    \label{BlackMirrorSchwarzschild}
    ds^{2}=-\frac{2m}{r(\sigma)}\Big(\frac{\sigma}{4m}\Big)^{2}
    dv_{\pm}^{2}\pm\frac{\sigma}{2m}d\sigma dv_{\pm}
    +r(\sigma)^{2}d\Omega^{2}.
\end{equation}
Note that the ingoing coordinate $v_{+}$ smoothly extends exterior $I$ onto the upper BH, and exterior $I'$ onto the lower BH (Fig.~\ref{stationary-mirror}a); and for this extension to be well defined (to not assign the same $(v_+,\sigma)$ coordinate to two distinct points) we must identify antipodal points on the upper and lower BH horizons, as shown in Fig.~\ref{stationary-mirror}c ({\it e.g.} $P\sim P', Q\sim Q'$).  Similarly, the ingoing coordinate $v_{-}$ extends $I$ onto the lower WH and $I'$ onto the upper WH (Fig.~\ref{stationary-mirror}b), and we must identify antipodal points on the two WH horizons ({\it e.g.} $R\sim R', S\sim S'$).\footnote{Note that we do {\it not} identify {\it all} antipodal points in Fig.~\ref{fig-kruskal-1}a, as has sometimes been considered in the past \cite{Gibbons:1986rzu}; in particular, we do not identify points in exterior $I$ with points in exterior $I'$.}  So on the ``other side" of the BH horizon from exterior $I$ is exterior $I'$ (rather than interior $II$); and similarly for the WH horizon.  This has glued exterior $I$ directly to exterior $I'$, and removed interiors $II$ and $II'$ altogether (Fig.~\ref{stationary-mirror}c).

The resulting spacetime is the simplest (Schwarzschild) black mirror: eternal, static, uncharged.  It is described by the metric (\ref{BlackMirrorSchwarzschild}), and covered by two coordinate patches, which together form a complete atlas: the ingoing patch ($-\infty<v_+<\infty, -\infty<\sigma<\infty$) which covers everything except the WH horizon, and the outgoing patch ($-\infty<v_-<\infty, -\infty<\sigma<\infty$) which covers everything except the BH horizon.
In these coordinates, all components of the metric $g_{\mu\nu}$ and Riemann tensor $R^{\alpha}_{\;\;\beta\gamma\delta}$ (and all curvature invariants, such as the Kretschmann scalar) are everywhere smooth, analytic and finite, and the vacuum Einstein equations $R_{\mu\nu}=0$ are satisfied.

Note that the Lorentzian and Euclidean geometries are now in perfect correspondence: in both, the $\sigma=0$ surface ({\it i.e.}\ the origin of the double cigar in Euclidean signature, or the event horizon in Lorentzian signature) serves as the boundary between two mirror-image {\it exterior} regions, with no interior regions or curvature singularities.

But there is no free lunch!  Although we have gotten rid of the black hole's curvature singularity, it has been replaced by another (milder) singularity.  In particular, if we compute the eigenvalues of $g_{\mu\nu}$ (in $v,\sigma$ coordinates) then, in addition to the angular eigenvalues $\lambda_{\theta}=r^{2}(\sigma)$ and $\lambda_{\varphi}=r^{2}(\sigma){\rm sin}^{2}\theta$, we see the remaining two (positive and negative) eigenvalues
\begin{equation}
  \label{sigma_pm}
  \lambda_{\pm}=-\frac{\sigma}{4r(\sigma)}
  \left(\frac{\sigma}{4m}\pm\sqrt{\frac{r(\sigma)^{2}\!\!}{m^{2}}+\frac{\sigma^{2}}{16m^{2}}}\right)
\end{equation}
swap places across the horizon, $\lambda_{+}(\sigma)=\lambda_{-}(-\sigma)$; so $\lambda_{\pm}(\sigma)$ both have simple analytic zeros at $\sigma=0$ (while the corresponding eigenvalues $\lambda_{\pm}^{-1}$ of $g^{\mu\nu}$ have simple poles). 

\begin{figure}
\begin{center}
\includegraphics[height=2.4in,width=2.4in]{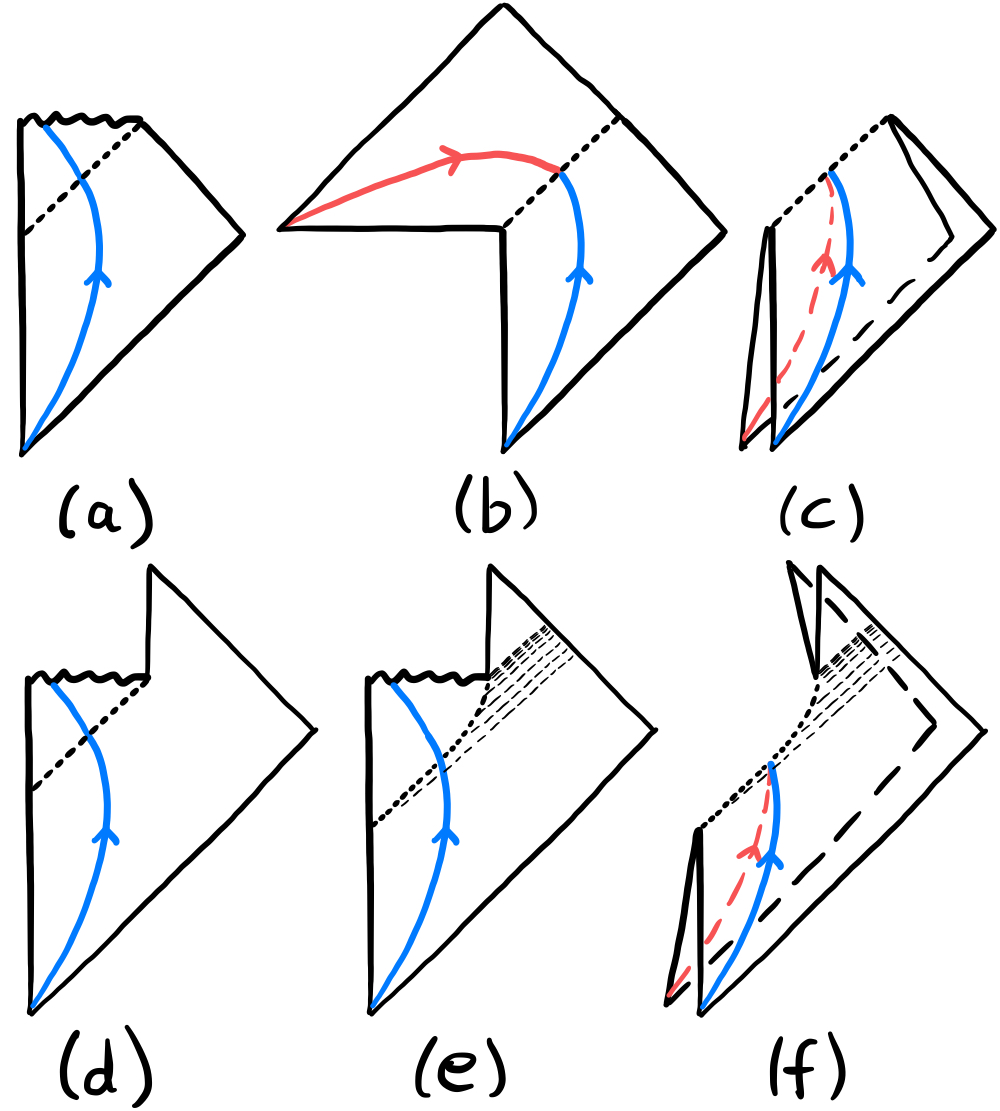}
\caption{The Penrose diagram for: a black hole formed by gravitational collapse (a); a black mirror formed by gravitational collapse, in unfolded view (b) or folded view (c); a black hole which collapses, then evaporates (d), improved to show shrinking of horizon due to Hawking evaporation (e); a black mirror which collapses, then evaporates (f).  Some infalling particle trajectories are also shown, see Appendices \ref{CPT_appendix}, \ref{geodesics}.
}
\label{collapsed-mirror}
\end{center}
\end{figure}

In the standard black hole solution, we are used to the fact that, if some eigenvalues of $g_{\mu\nu}$ approach zero or infinity as we approach the horizon, this is a mere coordinate singularity that can be removed by an appropriate change of coordinates.  But in the black mirror, this is not possible -- in coordinates where the metric is smooth and continuous across the horizon, the simple zeros in two eigenvalues of $g_{\mu\nu}$ are unavoidable: they represent a genuine (but mild) singularity in the geometry, closely related to the singularity joining the two sides of the Euclidean Schwarzschild spacetime in Fig.~\ref{euclidean-bh}b; and the corresponding simple poles in $g^{\mu\nu}$ represent real phenomena, akin to the poles we go around in defining the Feynman propagator.  We will explain their meaning below.

But first note that, although we have so far focused for simplicity on the Schwarzschild ({\it i.e.}\, 4D non-rotating, non-charged, asymptotically flat) black mirror, all the analysis and discussion extends directly -- without any qualitative changes -- to the most general 4D {\it stationary} black mirror (including rotation, electric and magnetic charge, and a non-zero cosmological constant).  This is explained in detail in Appendix \ref{BlackMirrorGeneral}.  
One can similarly obtain the spinning (A)dS black mirror in arbitrary dimensions, starting from the solution in \cite{Gibbons:2004js, Gibbons:2004uw}.

\section{Discussion}

\begin{figure}
\begin{center}
\includegraphics[height=1.4in,width=2.8in]{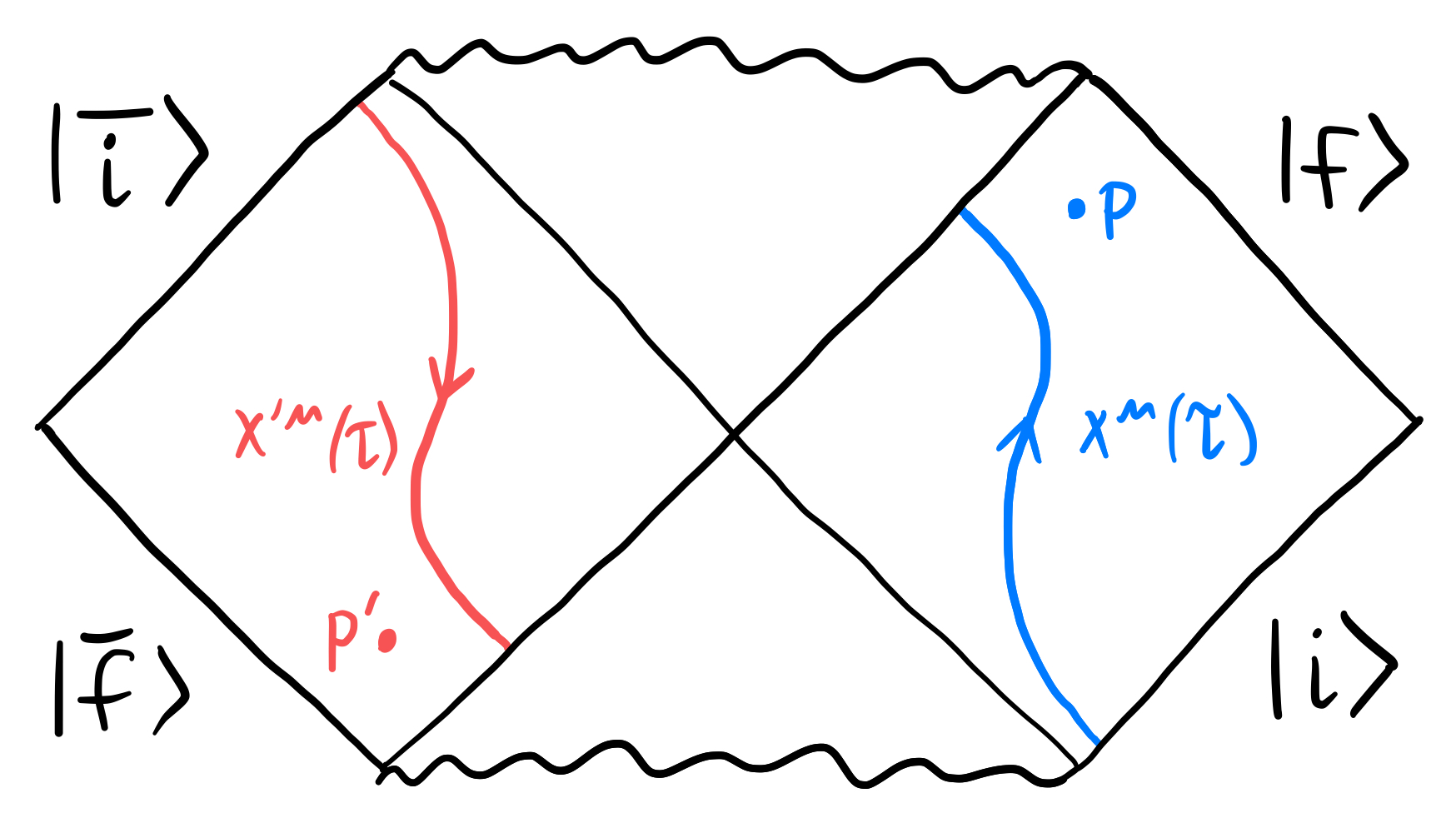}
\caption{Asymptotically flat spacetime with CPT-symmetric boundary conditions: states $|\bar{i}\rangle$ and $|\bar{f}\rangle$ are the CPT conjugates of states $|i\rangle$ and $f\rangle$, respectively (see Appendix \ref{CPT_appendix}).}
\label{chargeflow}
\end{center}
\end{figure}

So far, we have considered the general stationary black mirror.  Now let us explain how we think the general time-dependent case works.  (Since this part of the story can no longer be described explicitly analytically, it is correspondingly less certain and more speculative.)  We start with the Penrose diagram in Fig.~\ref{collapsed-mirror}a for an asymptotically-flat black hole formed by realistic gravitational collapse, with a blue curve showing the trajectory of an infalling particle (the arrow shows the direction of flow of electric charge).  The corresponding black mirror is again obtained by identifying two CPT mirror copies of the exterior at their respective horizons, as in Fig.~\ref{collapsed-mirror}b: now we see the infalling blue trajectory's smooth analytic extension (the red trajectory, see Appendix \ref{geodesics}).
For visualization, it is convenient to fold the diagram in Fig.~\ref{collapsed-mirror}b to obtain the diagram shown in Fig.~\ref{collapsed-mirror}c.  In this folded picture, time runs in its usual upward direction on both sheets, so we can explicitly see that both sheets share a common {\it future} null boundary ({\it i.e.}, a BH), and the previous trajectory may now be reinterpreted as the (blue) particle falling into the BH from one side annihilating with its CPT mirror (red) anti-particle falling in from the other. (The arrow flip is  explained in Appendix \ref{CPT_appendix}.)

In the stationary case, the event horizon and apparent horizon coincide.  More generally, where they do not, we conjecture the correct matching surface is the {\it apparent horizon} (which is well-defined to linear order in perturbations around the stationary solution \cite{Hollands:2024vbe}) since: (i) it is defined locally (unlike the event horizon); and (ii) it appears to be the surface associated to dynamical black hole entropy \cite{Hollands:2024vbe} which, as argued above, is also associated to the surface separating the two black mirror exteriors. 

Finally, some further hints favoring black mirrors:

i) It is natural to consider the path integral for our universe with CPT-symmetric boundary conditions as shown in Fig.~\ref{chargeflow} (see Appendices \ref{CPT_appendix}, \ref{saddle-point-app}).  Whereas the black hole (Fig.~\ref{fig-kruskal-1}a) and black mirror (Fig.~\ref{fig-kruskal-1}b) both satisfy these boundary condition, the black mirror seems to be the relevant saddle point as it (i) is free of curvature singularities, and (ii) has no {\it additional} boundaries (whereas the black hole solution has additional boundaries which are singular and also require further data to be specified).

ii) According to an external observer, an infalling object appears to take infinitely long to reach the BH.  In the standard account \cite{Oppenheimer:1939ue}, this is an optical illusion: the infalling object actually crosses the horizon into the interior, but it takes infinitely long for the visual evidence to reach the exterior observer.  However, taking quantum effects into account, the external observer sees the BH evaporate in a {\it finite} time, {\it before} the infalling object reaches the horizon, contradicting the standard account and drawing the interior's existence into question \cite{Vachaspati:2006ki, Vachaspati:2007hr, Kawai:2013mda}.

iii) The black hole information paradox \cite{Hawking:1976ra} and firewall paradox \cite{Almheiri:2012rt} come from imagining that quantum information falls into a black hole, and asking how it can possibly re-emerge (since the black hole interior is causally disconnected from future null infinity, see Figs.~\ref{collapsed-mirror}a, \ref{collapsed-mirror}d).  
The black mirror, with no interior, suggests a resolution: when a (blue) particle and (red) anti-particle meet at the horizon, they annihilate and the information 
is then confined to the horizon (classically, Figs.~\ref{collapsed-mirror}b, \ref{collapsed-mirror}c), or gradually leaks off the horizon (quantumly), until the black hole evaporates (Fig.~\ref{collapsed-mirror}f).

\begin{figure}
\begin{center}
\includegraphics[height=1.4in,width=2.8in]{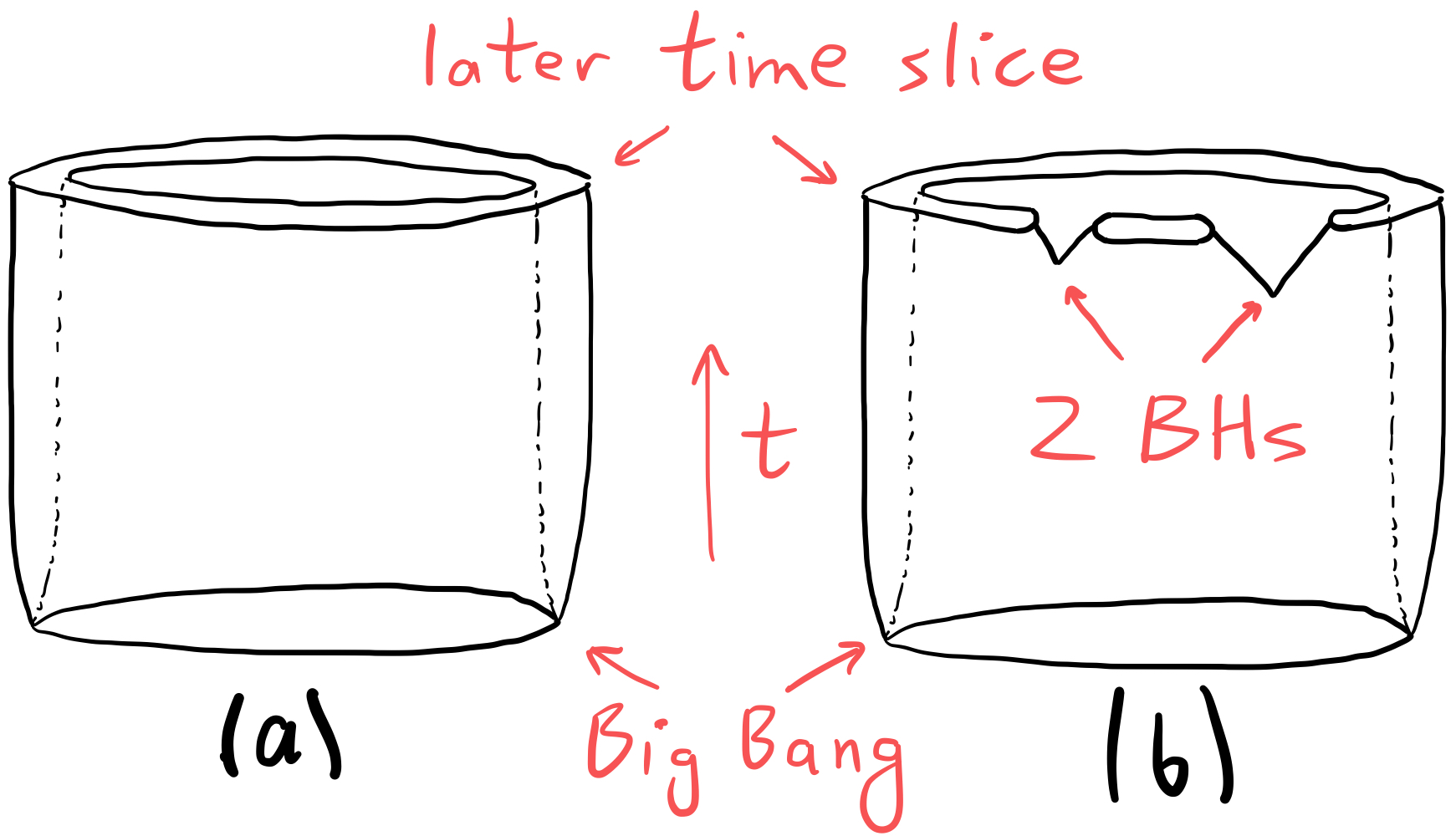}
\caption{The Penrose diagram for: (a) an unperturbed CPT-symmetric (two-sheeted) universe; (b) a perturbed CPT-symmetric universe in which two black mirrors have formed.}
\label{cosmology}
\end{center}
\end{figure}

iv) As a black hole settles to stationarity, we {\it expect} it to approach a thermal state, which should also be CPT-symmetric \cite{Sewell:1982zz}.  The black hole (Fig.~\ref{collapsed-mirror}a) does not do this (it settles down to a spacetime with a BH and no WH, with no way for CPT to act), but the black mirror (Fig.~\ref{collapsed-mirror}b, \ref{collapsed-mirror}e) {\it does} (see Appendix \ref{CPT_appendix}).  (Hawking also argued that CPT conflicts with black hole interiors \cite{Hawking:2014tga}.)

v) In the standard picture, one can drop global net charge into a black hole, which is then erased when the hole evaporates, leading to the lore that quantum gravity violates global symmetries \cite{Hawking:1975vcx, Witten:2017hdv, Harlow:2018jwu, Harlow:2018tng}.  By contrast, in the black mirror, global charge falling in on one side is canceled by anti-charge falling in on the other; so 
evaporation does {\it not} imply violation of global symmetry. 

vi) This CPT-symmetric picture for black mirrors fits naturally in the CPT-symmetric picture of the cosmos advocated in \cite{Boyle:2018tzc, Boyle:2018rgh, Boyle:2021jej, Boyle:2021jaz, Boyle:2022lyw, Turok:2022fgq, Boyle:2022lcq, Turok:2023amx} (and \cite{Kumar:2024nhe}) (see Fig.~\ref{cosmology}): the two mirror-image sheets of spacetime split at the Bang, and merge at BHs.  As before, it is natural to regard the two-sheeted spacetime as representing a pure state, while one sheet represents the mixed state obtained by tracing over the other sheet.  This neatly explains why the corresponding entropy, associated with the boundary surface separating the two sheets, is given by the {\it horizon area} (in the BH case) \cite{Bekenstein:1972tm, Bekenstein:1973ur, Gibbons:1976ue}, and the {\it total comoving spatial volume} (in the cosmological case \cite{Turok:2022fgq, Boyle:2022lcq}).

\begin{acknowledgments}
KT wishes to thank Mihalis Dafermos and Matt Visser for their helpful comments during the Strong Gravity conference.  LB is grateful to Minhyong Kim for helpful discussions.  KT, LB, and NT are supported by the STFC Consolidated Grant `Particle Physics at the Higgs Centre', and NT is supported by the Higgs Chair at the University of Edinburgh. Research at Perimeter Institute is supported by the Government of Canada, through Innovation, Science and Economic Development, Canada and by the Province of Ontario through the Ministry of Research, Innovation and Science. 
\end{acknowledgments}

\bibliography{ref.bib}

\newcommand{\modone}{~(\text{mod} 1)}
\begin{appendix}

\section{General Stationary black mirror}
\label{BlackMirrorGeneral}

\subsection{Boyer-Lindquist coordinates}

Consider the general charged, rotating black hole exterior in dS or AdS \cite{Carter:1968ks,Carter:1973rla, Plebanski:1976gy, Caldarelli:1999xj}.  We start in generalized Boyer-Lindquist coordinates $(t,r,\theta,\phi)$ -- the generalization of the Schwarzschild coordinates considered Section \ref{Schwarzschild}. We define the tetrad
\begin{subequations}
  \begin{eqnarray}
    e^{0}&=&\frac{\Delta_{r}^{1/2}}{\rho}(dt-\frac{a\,{\rm sin}^{2}\theta}{\Xi}d\phi) \\
    e^{1}&=&\frac{\rho}{\Delta_{r}^{1/2}}dr \\
    e^{2}&=&\frac{\rho}{\Delta_{\theta}^{1/2}}d\theta \\
    e^{3}&=&\frac{\Delta_{\theta}^{1/2}{\rm sin}\,\theta}{\rho}(a\,dt-\frac{r^{2}+a^{2}}{\Xi}d\phi)
  \end{eqnarray}
\end{subequations}
where we have defined
\begin{eqnarray}
  \Delta_{\theta}\equiv 1\!-\!\frac{a^{2}}{\ell^{2}}{\rm cos}^{2}\theta,&\!\!&
  \Delta_{r}\equiv(r^{2}\!+\!a^{2})\Big(1\!+\!\frac{r^{2}}{\ell^2}\Big)\!-\!2mr\!+\!q^{2},\nonumber\\
  \rho^{2}\equiv r^{2}\!+\!a^{2}{\rm cos}^{2}\theta,&\!\!& 
  \Xi\equiv 1-\frac{a^{2}}{\ell^2}, 
\end{eqnarray}
with $a$ the rotation parameter, $\ell$ the AdS radius, and $q^{2}=q_{e}^{2}+q_{m}^{2}$,  with $q_{e}$ and $q_{m}$ the electric and magnetic charges, respectively.  (To obtain the 
corresponding dS black hole, just make the substitution $\ell^2\to-\ell^2$.) Then the line element is
\begin{subequations}
\begin{equation}
    ds^{2}=\eta_{ab}e^{a}e^{b}\qquad\qquad \eta_{ab}={\rm diag}(-1,1,1,1)
\end{equation}
and the electromagnetic potential $A=A_{\mu}dx^{\mu}$ and field strength $F=dA=F_{\mu\nu}dx^{\mu}\wedge dx^{\nu}$ are given by
\begin{equation}
  A=-\frac{q_{e}r}{\Delta_{r}^{1/2}\rho}e^{0}-\frac{q_{m}{\rm cos}\,\theta}{\Delta_{\theta}^{1/2}\rho\,{\rm sin}\,\theta}e^{3}
\end{equation}
and
\begin{eqnarray}
  \label{F_BoyerLindquist}
  F\!&\!=\!&\!-\frac{1}{\rho^{4}}[\;\!q_{\;\!e}(r^{2}\!-\!a^{2}{\rm cos}^{2}\theta)\!+\!2q_{m}r\;\!a\;\!{\rm cos}\;\!\theta]\,e^{0}\!\wedge\! e^{1} \nonumber\\
  &&+\frac{1}{\rho^{4}}[q_{m}(r^{2}\!-\!a^{2}{\rm cos}^{2}\theta)\!-\!2\;\!q_{\;\!e}r\;\!a\;\!{\rm cos}\;\!\theta]\,e^{2}\!\wedge\! e^{3}.\qquad
\end{eqnarray}
\end{subequations}
Note that if $m<m_{extr}$, where $m_{extr}$ is defined in Eq.~(8) in Caldarelli, Cognola and Klemm (1999), then the black hole is super-extremal; $\Delta_{r}$ has no real roots and hence no horizons, and a naked singularity.  We will henceforth assume the opposite, that $m>m_{extr}$, so that the black hole is sub-extremal, with two real roots, $r_{-}$ and $r_{+}$.  The BH horizon is located at the outer root $r_{+}$, and the horizon area is given by
\begin{equation}
  {\cal A}_{BH}=\frac{4\pi(r_{+}^{2}+a^{2})}{\Xi}.
\end{equation}

Boyer-Lindquist coordinates make manifest the full symmetry of the exterior spacetime: two Killing vectors $\partial/\partial t$ and $\partial/\partial\varphi$, and two discrete symmetries which send $(t,r,\theta,\phi)$ to, respectively, $(t,r,\pi-\theta,\phi)$ and $(-t,r,\theta,-\phi)$.

\subsection{Going Euclidean, thermodynamics}

To construct the Euclidean metric, we start with Boyer-Lindquist coordinates $(t,r,\theta,\phi)$, and make the Wick rotations $t=i\tau$, $a=i\hat{a}$, $q=i\hat{q}$, and consider real values of $\tau$, $\hat{a}$, and $\hat{q}$, thus obtaining the new tetrad
\begin{subequations}
  \begin{eqnarray}
    \hat{e}^{0}&=&\frac{\hat{\Delta}_{r}^{1/2}}{\hat{\rho}}(d\tau-\frac{\hat{a}\,{\rm sin}^{2}\theta}{\hat{\Xi}}d\phi) \\
    \hat{e}^{1}&=&\frac{\hat{\rho}}{\hat{\Delta}_{r}^{1/2}}dr \\
    \hat{e}^{2}&=&\frac{\hat{\rho}}{\hat{\Delta}_{\theta}^{1/2}}d\theta \\
    \hat{e}^{3}&=&\frac{\hat{\Delta}_{\theta}^{1/2}{\rm sin}\,\theta}{\hat{\rho}}(-\hat{a}\,d\tau-\frac{r^{2}-\hat{a}^{2}}{\hat{\Xi}}d\phi)
  \end{eqnarray}
\end{subequations}
where
\begin{eqnarray}
  \hat{\Delta}_{\theta}\equiv 1\!+\!\frac{\hat{a}^{2}}{\ell^{2}}{\rm cos}^{2}\theta,&\!\!&\hat{\Delta}_{r}\equiv(r^{2}\!-\!\hat{a}^{2})\Big(1\!+\!\frac{r^{2}}{\ell^2}\Big)\!-\!2mr\!-\!\hat{q}^{2}, \nonumber\\
  \hat{\rho}^{2}\equiv r^{2}\!-\!\hat{a}^{2}{\rm cos}^{2}\theta,&\!\!&\hat{\Xi}=1+\frac{\hat{a}^{2}}{\ell^2}.
\end{eqnarray}
Then the Euclidean line element is
\begin{equation}
  ds^{2}=\delta_{ab}\hat{e}^{a}\hat{e}^{b}
\end{equation}
Next we switch radial coordinate $r\to\sigma$, by making the substitution
\begin{equation}
  \label{r(sigma)_gen}
  r(\sigma)\to\hat{r}_{+}+\alpha^{2}\sigma^2
  \quad\Rightarrow\quad
  dr\to 2\alpha^{2}\sigma d\sigma
\end{equation}
where $\hat{r}_{+}$ is the Wick rotation of the previous horizon radius $r_{+}$ (and a root of $\hat{\Delta}_{r}$), the constant $\alpha$ is determined below.  Then, to leading order in $\sigma$, the metric becomes
\begin{equation}
  \label{StationaryBlackMirror}
  ds^{2}=\frac{4\alpha^{2}\hat{\rho}_{+}^{2}}{\hat{\Delta}_{r}'(\hat{r}_{+})}(d\sigma^{2}+\sigma^{2}d\hat{\omega}_{1}^{2})+\frac{\hat{\rho}_{+}^{2}}{\hat{\Delta}_{\theta}}(d\theta^{2}+{\rm sin}^{2}\theta\,d\hat{\omega}_{2}^{2})
\end{equation}
where
\begin{subequations}
  \begin{eqnarray}
    d\hat{\omega}_1&=&\frac{\hat{\Delta}_{r}'(\hat{r}_{+})}{2\hat{\rho}_{+}^{2}}(d\tau-\frac{\hat{a}\,{\rm sin}^{2}\theta}{\hat{\Xi}}d\phi) \\
    d\hat{\omega}_2&=&\frac{\hat{\Delta}_{\theta}}{\hat{\rho}_{+}^{2}}(\hat{a}d\tau+\frac{r_{+}^{2}-\hat{a}^{2}}{\hat{\Xi}}d\phi)
  \end{eqnarray}
\end{subequations}
and we see that, by choosing $\alpha^{2}=\hat{\Delta}_{r}'(\hat{r}_{+})/(4\hat{\rho}_{+}^{2})$ we can set the leading coefficient in (\ref{StationaryBlackMirror}) 
to unity.

As in the Schwarzschild case, we see the Euclidean metric has the topology $C^{2}\times S^{2}$ of a one-sided 2-cone $C^{2}$  (parameterized by $\tau$, $\sigma$, with its tip at $\sigma=0$) times a 2-sphere $S^2$ (parameterized by $\theta$, $\varphi$) -- {\it i.e.}\ the blue side of Fig.~\ref{euclidean-bh}a. And, as before, we can analytically extend this to a 2-sided cone by letting $\sigma$ extend to negative values (the blue+red surface in Fig.~\ref{euclidean-bh}a).  Note that this extended metric now has a new $\mathbb{Z}_{2}$ isometry $\sigma\to-\sigma$.  

In order to make the metric smooth at the cone's tip (to remove the conic singularity, as in Fig.~\ref{euclidean-bh}b), we must identify $\hat{\omega}_{1}\sim \hat{\omega}_{1}+2\pi$; and, in since this identification must be $\theta$-independent, we are forced to identify
\begin{equation}
  (\tau,\phi)\;\sim\;(\tau+2\pi\hat{\beta},\varphi+2\pi i\hat{\Omega}\hat{\beta})
\end{equation}
where
\begin{subequations}
  \begin{eqnarray}
    \hat{\beta}&=&\frac{2(\hat{r}_{+}^{2}-\hat{a}^{2})}{\hat{\Delta}_{r}'(\hat{r}_{+})} \\
    \hat{\Omega}&=&\frac{i\hat{a}\hat{\Xi}}{\hat{r}_{+}^{2}-\hat{a}^{2}}.
  \end{eqnarray}
\end{subequations}
Note that this same shift leaves $\omega_{2}$ invariant.

Now, Wick-rotating back to Lorentzian signature, and expressing everything in terms of our original Lorentzian variables, we have the identification
\begin{equation}
  (t,\phi)\;\sim\;(t+2\pi i\beta,\varphi+2\pi i\Omega\beta)
\end{equation}
where
\begin{subequations}
  \begin{eqnarray}
    \beta&=&\frac{2(r_{+}^{2}+a^{2})}{\Delta_{r}'(r_{+})}, \\
    \Omega&=&\frac{a\Xi}{r_{+}^{2}+a^{2}}.
  \end{eqnarray}
\end{subequations}
Here $\beta$ is identified as the temperature of the black hole, and $\Omega$ as the angular velocity of the horizon.

\subsection{In Eddington-Finkelstein coordinates}

Although the Boyer-Lindquist system has the advantage that it makes manifest the full symmetry of the metric, it has the disadvantage that it only covers the exterior $r>r_{+}$, and fails to extend to (or beyond) the black and white horizons (where it becomes singular).  To fix this, we first define new generalized ingoing/outgoing Eddington-Finkelstein coordinates $(v_{\pm},r,\theta,\hat{\phi})$ that remain non-singular on one horizon or the other, by a coordinate transformation of the form
\begin{subequations}
  \begin{eqnarray}
    dv&=&dt+\alpha_t\,dr \\
    d\hat{\phi}&=&d\phi+\alpha_{\phi}\,dr
  \end{eqnarray}
\end{subequations}
where $\alpha_t=\alpha_t(r)$ and $\alpha_{\phi}=\alpha_{\phi}(r)$ are chosen to make the $dr^2$ term drop out of the line element $ds^2$ in the new (Eddington-Finkelstein) coordinate system.  
One can check that this implies
\begin{subequations}
  \begin{eqnarray}
    \alpha_t&\equiv&\pm\frac{r^{2}+a^{2}}{\Delta_{r}} \\
    \alpha_{\phi}&\equiv&\pm\frac{a\,\Xi}{\Delta_{r}}
  \end{eqnarray}
\end{subequations}
where the $+/-$ sign corresponds to ingoing/outgoing Eddington-Finkelstein coordinates which cover the exterior and black/white horizon, respectively.

In these new coordinates, if we define the new tetrad
\begin{subequations}
  \label{StationaryBlackMirrorTetrad}
  \begin{eqnarray}
    \tilde{e}^{0}&=&dv-\frac{a\,{\rm sin}^{2}\theta}{\Xi}d\hat{\phi} \\
    \tilde{e}^{1}&=&dr \\
    \tilde{e}^{2}&=&\frac{\rho}{\Delta_{\theta}^{1/2}}d\theta \\
    \tilde{e}^{3}&=&\frac{\Delta_{\theta}^{1/2}{\rm sin}\,\theta}{\rho}(a\,dv-\frac{r^{2}+a^{2}}{\Xi}d\hat{\phi})
  \end{eqnarray}
\end{subequations}
the line element is
\begin{subequations}
  \label{StationaryBlackMirrorMetric}
  \begin{eqnarray}
    ds^{2}&=&\tilde{\eta}_{ab}\tilde{e}^{a}\tilde{e}^{b},
    \quad\tilde{\eta}_{ab}=\left(\begin{array}{cccc}
    -\frac{\Delta_{r}}{\rho^{2}} & \pm 1 & 0 & 0 \\
    \pm 1 & 0 & 0 & 0 \\
    0 & 0 & 1 & 0 \\
    0 & 0 & 0 & 1 \end{array}\right)\quad
  \end{eqnarray}
  the gauge potential is
  \begin{equation}
    \label{StationaryBlackMirrorA}
    A=-\frac{q_{e}r}{\rho^2}\tilde{e}^{0}-\frac{q_{m}{\rm cos}\,\theta}{\Delta_{\theta}^{1/2}\rho\,{\rm sin}\,\theta}\tilde{e}^{3}
  \end{equation}
  (where we removed an additional term $\pm\frac{q_{e}r}{\Delta_{r}}dr$ via a gauge transformation), and the field strength is
  \begin{eqnarray}
    \label{StationaryBlackMirrorF}
    F\!&\!=\!&\!-\frac{1}{\rho^{4}}[\;\!q_{\;\!e}
    (r^{2}\!-\!a^{2}{\rm cos}^{2}\theta)\!+\!2q_{m}r\;\!a\;\!{\rm cos}\;\!\theta]\,\tilde{e}^{0}\!\wedge\!\tilde{e}^{1} \nonumber\\
    &&+\frac{1}{\rho^{4}}[q_{m}(r^{2}\!-\!a^{2}{\rm cos}^{2}\theta)\!-\!2\;\!q_{\;\!e}r\;\!a\;\!{\rm cos}\;\!\theta]\,\tilde{e}^{2}\!\wedge\! \tilde{e}^{3}.\qquad\;\;
  \end{eqnarray}
\end{subequations}
We see that these coordinates remain non-singular at the black/white horizon $r=r_{+}$ (where $\Delta_{r}$ vanishes): the components of $g_{\mu\nu}$, $A_{\mu}$ and $F_{\mu\nu}$ all remain finite, and the metric determinant remains finite and non-zero.  

Now, as before, we can switch from $r\to\sigma$ by making the substitution (\ref{r(sigma)_gen}) in Eqs.~(\ref{StationaryBlackMirrorTetrad}, \ref{StationaryBlackMirrorMetric}, \ref{StationaryBlackMirrorA}, \ref{StationaryBlackMirrorF}) and extending $\sigma$ to negative values.  If we now let the coordinates $v_{+}$ and $\sigma$ (respectively, $v_{-}$ and $\sigma$) extend from $-\infty$ to $+\infty$, we obtain the general stationary black mirror, described in ingoing coordinates $(v_{+},\sigma,\theta,\hat{\varphi})$ (respectively, outgoing coordinates $(v_{-},\sigma,\theta,\hat{\varphi})$) which cover the entire spacetime except for the black horizon (respectively, white horizon).  Taken together, these ingoing and outgoing coordinate systems provide an atlas covering the complete stationary black mirror spacetime.  

As before, the extended spacetime now has a new isometry $\sigma\to-\sigma$ that takes us back and forth between the two exterior regions ($I$ and $I$'), which are identified at the horizon, as in Fig.~\ref{fig-kruskal-1}b.

Note that, in these $(v_{\pm},\sigma,\theta,\hat{\varphi})$ coordinates, the components of the resulting metric $g_{\mu\nu}$ and field strength $F_{\mu\nu}$, as well as the components of all curvature invariants (such as the Kretschmann scalar $R_{\alpha\beta\gamma\delta}R^{\alpha\beta\gamma\delta}$) are again everywhere explicitly smooth, analytic and finite, so the spacetime has no curvature singularities.  But, as in the Schwarzschild case, two of the eigenvalues $\lambda_{\pm}$ of the metric $g_{\mu\nu}$ are exchanged as we cross the horizon, $\lambda_{+}(\sigma)=\lambda_{-}(-\sigma)$, and both of these eigenvalues have simple analytic zeros at $\sigma=0$, while the corresponding eigenvalues $\lambda_{\pm}^{-1}$ of $g^{\mu\nu}$ have simple poles).

\section{CPT symmetry}
\label{CPT_appendix}

In the body of the paper, we focused for simplicity on the {\it uncharged} (Schwarzschild) black mirror but, to understand the action of $CPT$ more clearly, let us now consider the {\it charged} (Reissner-Nordstrom) black mirror. The natural $CPT$ transformation of the black hole background (the unique involutive isometry which commutes with all of the continuous isometries, and has no fixed points) \cite{Gibbons:1986rzu} sends 
\begin{equation}
    (t,\sigma,\theta,\varphi)\to(t,-\sigma,\pi-\theta,\pi+\varphi).
\end{equation}
This isometry maps {\it e.g.} a point $P$ near the upper corner of exterior $I$ (and with the angular coordinates $\theta,\varphi$) to a mirror point $P'$ near the lower corner of exterior $I'$ (with antipodal angular coordinates $\pi-\theta,\pi+\varphi$); see Fig. \ref{chargeflow}.  

The black mirror is obtained from the corresponding black hole by identifying each point on the BH (WH) horizon bounding exterior $I$ with its CPT image on the BH (WH) horizon bounding exterior $I'$.

\subsection{Orientations and $CPT$ versus $PT$}

In this subsection, let us explain why the isometry $\varphi_{CPT}: (t,\sigma,\theta,\varphi)\to(t,-\sigma,\pi-\theta,\pi+\varphi)$ (which inverts the angular 2-sphere) is a classical analogue of the $CPT$ transformation, whereas the isometry $\varphi_{PT}: (t,\sigma,\theta,\varphi)\to(t,-\sigma,\theta,\varphi)$ (which does {\it not} invert the angular 2-sphere) corresponds to $PT$.

\begin{figure}
\begin{center}
\includegraphics[height=2.0in,width=2.4in]{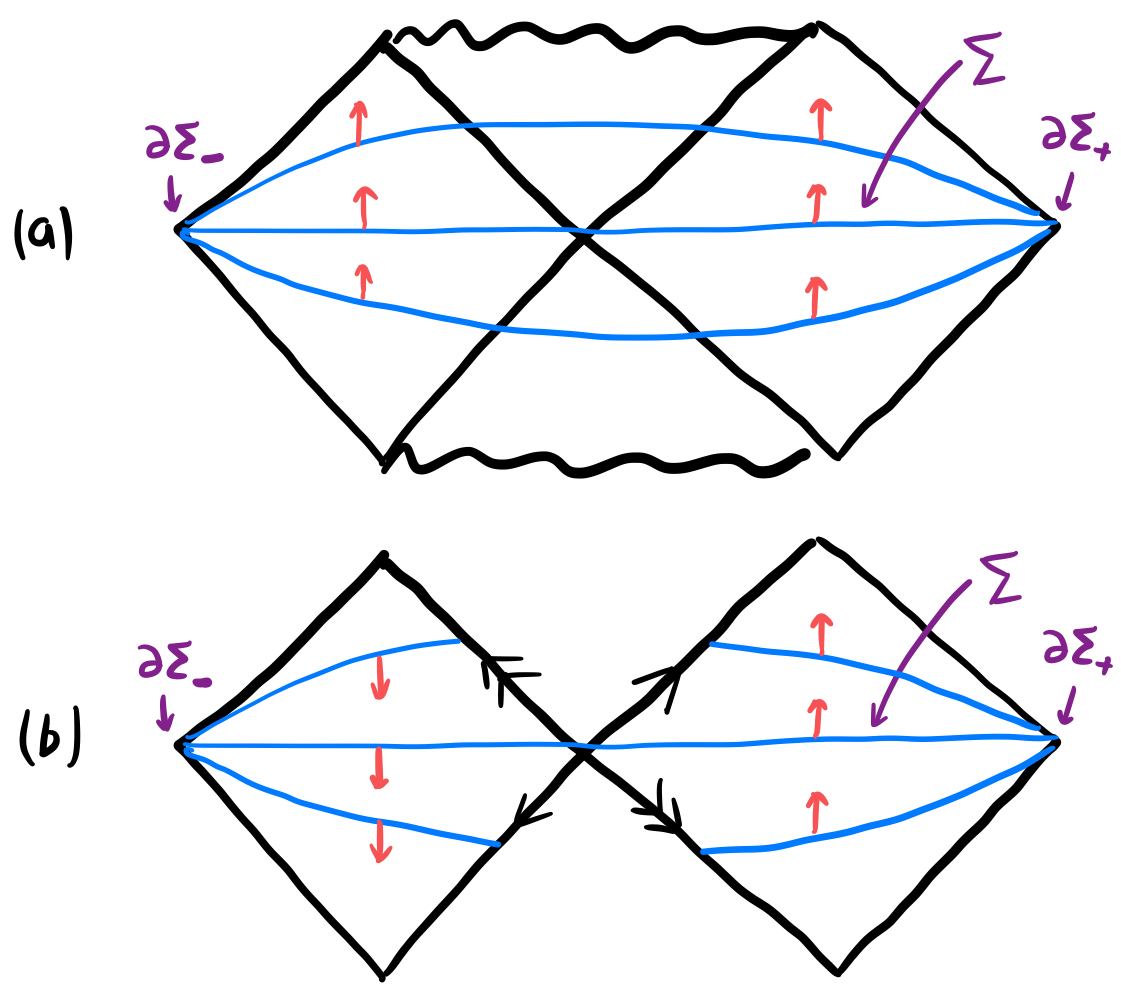}
\caption{The time orientation in (a) a black hole and
(b) a black mirror.}
\label{BlackMirrorOrientation}
\end{center}
\end{figure}

We start by reviewing the space, time, and spacetime orientations of the maximally-extended Schwarzschild black hole.  In null Kruskal coordinates $(U,V,\theta,\varphi)$, the line element is
\begin{equation}
  ds^{2}=-\frac{32m^{3}}{r}{\rm e}^{-r/2m}dU dV+r^{2}d\Omega^{2}
\end{equation}
with $r=r(U,V)$ given by
\begin{equation}
  r=2m\left[1+W_{0}\left(-\frac{UV}{e}\right)\right]
\end{equation}
where $W_{0}$ is the Lambert $W$ function, and $U$ and $V$ both run from $-\infty$ to $+\infty$.  This spacetime has a global 4-orientation, {\it i.e.}, an everywhere non-vanishing 4-form $\mu_{4}$ (the spacetime volume 4-form).   In terms of the above coordinates it is given by 
\begin{equation}
  \mu_{4}=\frac{16m^{3}}{r}{\rm e}^{-r/2m}r^{2}{\rm sin}\,\theta\,dU\wedge dV\wedge d\theta\wedge d\varphi.
\end{equation}
To determine whether an ordered tetrad of vectors is positively or negatively oriented, we contract it with $\mu_{4}$ and check the sign of the result. In this way, we see that the tetrad of Kruskal coordinate vectors $(\partial_{U},\partial_{V},\partial_{\theta},\partial_{\varphi})$ is everywhere positively oriented (except at $\theta=0$ or $\pi$, where the spherical coordinates $\theta,\varphi$ on the 2-sphere have their usual coordinate singularity); and we can also check that the tetrad $(\partial_{t},\partial_{r},\partial_{\theta},\partial_{\varphi})$ is positively oriented in both exteriors, $I$ and $I'$.  

The maximally-extended Schwarschild black hole is foliated by global Cauchy surfaces and has a global time orientation -- {\it i.e.}\ an everywhere non-vanishing timelike vector field $T$ (which may be taken to consist of the timelike unit normal vectors to the previously-mentioned Cauchy surfaces).  For $r\gg2m$, $T$ agrees with $+\partial_t$ in exterior $I$, and $-\partial_t$ in exterior $I'$ (see Fig.~\ref{BlackMirrorOrientation}a).  The three-form $\mu_{3}$ obtained by contracting the time-orientation vector $T$ with the 4-orientation $\mu_{4}$ is then the 3-orientation on the black hole Cauchy surfaces. 

Under the $PT$ isometry $\varphi_{PT}$,  
one can check that: (i) the push-forward of the vector field $T$ is $-T$; (i) the pullback of the spatial orientation $\mu_{3}$ is $-\mu_{3}$; and (iii) the pullback of the spacetime orientation $\mu_{4}$ is $+\mu_{4}$.  In other words, $\varphi_{PT}$ 
inverts the time orientation and the spatial orientation, but preserves the spacetime orientation (as we would expect for $PT$).

Now, let us return to the more general (charged) black hole, and its electromagnetic field strength two-form $F$.  Using Gauss's law, we obtain the electric charge of the black hole (in one exterior) by integrating $\star F$ over an angular 2-sphere (in that exterior).  Under the $PT$ isometry $\varphi_{PT}$, the pullback of $\star F$ is $\star F$, whereas, under the $CPT$ isometry $\varphi_{CPT}$, the pullback of $\star F$ is $-(\star F)$.  
So $\varphi_{PT}$ preserves the hole's charge, while $\varphi_{CPT}$ reverses it (again, as we would expect for $PT$ and $CPT$).

\subsection{Total charge (and mass) of the black hole.}
\label{Black_Hole_Charge}

In the following subsection -- Appendix \ref{Black_Mirror_Charge} -- we will show that the two sides of the black {\it mirror} have the {\it same} electric charge and mass, as measured by Gauss's Law.

But first, in the present subsection, we review the corresponding result for the black {\it hole}: the two sides of the black {\it hole} have {\it opposite} electric charge (or mass), as measured by Gauss's Law, so that the total charge (or mass) of the maximally-extended (two-sided) black hole (again, as measured by Gauss's Law, and including the contributions from {\it both} exteriors) is {\it zero}.  

One way to obtain the black hole's total electric charge is to integrate the electric charge density over a complete Cauchy surface, {\it e.g.}\ the spatial 3-surface $\Sigma$ shown in Fig.~\ref{BlackMirrorOrientation}a.  But we are dealing with a vacuum solution where the charge density vanishes everywhere on this Cauchy surface, so the total electric charge must vanish as well!

We can obtain the same conclusion via Gauss's law.  In this approach, the charge is obtained by integrating $\star F$ over the Cauchy surface's spatial 2-boundary $\partial \Sigma$.  This boundary has two components: $\partial\Sigma_{+}$ (the $S^2$ at $r=\infty$ in region $I$) and $\partial\Sigma_{-}$ (the $S^2$ at $r=\infty$ in region $I'$).  When expressed in $(t,r,\theta,\varphi)$ coordinates, we see that $\star F\propto q\,{\rm sin}\,\theta\,d\theta\wedge d\varphi$ is identical in the two exteriors; but the two boundary components $\partial\Sigma_+$ and $\partial\Sigma_-$ have opposite orientations\footnote{To see that $\partial\Sigma_{+}$ and $\partial\Sigma_{-}$ have opposite orientations, note that, when expressed in terms of $(t,r,\theta,\varphi)$ coordinates, the 4-orientation $\mu_{4}=r^{2}{\rm sin}\,\theta\,dt\wedge dr\wedge d\theta\wedge d\varphi$ is identical in the two exteriors; and the orientation two-form $\mu_{2}$ on the 2-boundary $\partial\Sigma$ is obtained by contracting this 4-orientation $\mu_{4}$ with the outward pointing spatial normal vector $\partial_{r}$ (the same in both exteriors), and the time orientation vector $T$, which is $+\partial_{t}$ in exterior $I$ and $-\partial_{t}$ in exterior $I'$; so the two orientation $\mu_{2}$ is $\propto {\rm sin}\,\theta\,d\theta\wedge d\varphi$ in exterior $I$, and $\propto-{\rm sin}\,\theta\,d\theta\wedge d\varphi$ in exterior $I'$.} so the two integrals cancel.

A closely analogous argument shows that the total Komar mass of the maximally-extended (two-sided) black hole is zero.  Again, in this case, the argument can be expressed either in terms of integrating the mass density over the complete 3D Cauchy surface $\Sigma$, or by using Gauss's Law/Stokes' Theorem to rephrase this in terms of the Cauchy surface's two-component 2D boundary $\partial\Sigma=\partial\Sigma_{+}+\partial\Sigma_{-}$.  Consider the $2D$ ($\partial\Sigma$) version of the argument: just as we computed the total (2-sided) electric charge $Q_{tot}$ by starting with the potential 1-form $A=-(q_{e}/r)dt$, computing its field strength $F_{A}=dA$, and finding $Q_{tot}\propto \int_{\partial\Sigma_{+}}\star F_{A}+\int_{\partial\Sigma_{-}}\star F_{A}=0$, we can compute the total (2-sided) Komar mass $M_{tot}$ by starting with the Killing one-form $k=-(1-\frac{2m}{r})dt$ (the one-form dual to the timelike Killing vector $\partial_{t}$), computing its ``field strength" $F_{k}=dk$, and finding $M_{tot}\propto \int_{\partial\Sigma_{+}}\star F_{k}+\int_{\partial\Sigma_{-}}\star F_{k}=0$.

So, in this sense (with respect to its global time and spacetime orientation) the total charge (or mass) of the black hole vanishes; or, stated another way, its two exteriors have equal but opposite charge (or mass).

\subsection{Total charge (and mass) of the black mirror.}
\label{Black_Mirror_Charge}

Finally, in this section, we turn to the black mirror and find that something surprising happens, due to its non-trivial topology: as determined via Gauss's Law from the electric field (or gravitational field) flux through a distant two-sphere, the two sides now have the {\it same} charge (or mass), and hence a total two-sided charge (or mass) which does not cancel to zero (as it did in the two-sided black hole), but is rather {\it twice} that of an ordinary (one-sided) black hole; and this is true even though, contrary to Gauss's Law, the solution is a smooth (but topologically non-trivial) vacuum solution, with no ``enclosed" electric charge (or mass) localized on the mirror.  This is a novel incarnation of Misner and Wheeler's idea of ``mass without mass" and ``charge without charge" \cite{Misner:1957mt,Sorkin:1977wq,Giulini:2009ts}.

To understand this, let us denote the black mirror's spacetime 4-orientation by $\tilde{\mu}_{4}$ and its electromagnetic field strength by $\tilde{F}_{A}$. Note that, in the coordinates $\{v_{\pm},\sigma,\theta,\varphi\}$ or $\{t,\sigma,\theta,\varphi\}$ in which the mirror's two-sided metric $\tilde{g}_{\mu\nu}$ is smooth, its spacetime 4-orientation $\tilde{\mu}_{4}$ and electromagnetic field strength $\tilde{F}$ are also smooth: they are given by $\tilde{\mu}_{4}=2 r^{2}\sigma\,{\rm sin}\,\theta\,dt\wedge d\sigma\wedge d\theta\wedge d\varphi$ and $\tilde{F}_{A}=-2 (q \sigma/r^2)dt\wedge d\sigma$.  In fact, if we express the black hole quantities $\mu_{4}$, $F_{A}$ and $\star F_{A}$ in these same coordinates, we see that they agree with their black mirror counterparts $\tilde{\mu}_{4}$, $\tilde{F}_{A}$ and $\star{\tilde{F}}_{A}$ in the exterior regions $I$ and $I'$.  But note that the black mirror quantities $\tilde{\mu}_{4}$, $\tilde{F}_{A}$ and $\star{\tilde{F}}_{A}$ all vanish at the mirror itself ({\it i.e}\ at $\sigma=0$), a sign that something topologically non-trivial happens at the mirror, as we already knew from Fig.~\ref{euclidean-bh} and the metric eigenvalues $\lambda_{\pm}$ (\ref{sigma_pm}) smoothly swapping signs at $\sigma=0$.

Next note that the black mirror and black hole have {\it different} global time orientations.  This may be understood from Fig.~\ref{stationary-mirror}: from the standpoint of the black hole's global time orientation $T$ (see Fig.~\ref{BlackMirrorOrientation}a), point $P$ is to the future of point $Q$, whereas point $P'$ is to the past of point $Q'$.  But, since we identify $P\sim P'$ and $Q\sim Q'$ in the black mirror, in order for them to have a consistent time ordering, we learn that the black mirror must have a different global time orientation $\widetilde{T}$ (see Fig.~\ref{BlackMirrorOrientation}b).  In particular, the black mirror's time orientation $\widetilde{T}$ agrees with the black hole's time orientation $T$ in exterior $I$, but differs from $T$ by a sign in exterior $I'$.  

Now let us consider the 2-orientation on the black mirror's angular 2-boundary $\partial\Sigma=\partial\Sigma_{+}+\partial\Sigma_{-}$.  In the previous subsection, we computed the black hole's corresponding 2-orientation $\mu_{2}$ by contracting its spacetime 4-orientation $\mu_{4}$ with its time orientation $T$ and the outward pointing normal $\partial_{r}$.  Analogously, we now compute the black mirror's 2-orientation $\tilde{\mu}_{2}$ by contracting its spacetime 4-orientation $\tilde{\mu}_{4}$ with its time orientation $\tilde{T}$ and the outward pointing normal $\partial_{r}$.  Since $T$ and $\tilde{T}$ differ by a sign in exterior $I'$, $\mu_{2}$ and $\tilde{\mu}_{2}$ also differ by a sign in in exterior $I'$.  In other words, in the black mirror, $\star\tilde{F}$ is again $\propto+{\rm sin}\,\theta\,d\theta\wedge d\varphi$ on both sides, but now (in contrast to the black hole case) the mirror's angular 2-orientation is {\it also} $\propto+{\rm sin}\,\theta\,d\theta\wedge d\varphi$ on both sides.  

So if we use Gauss's law to determine the mirror's total two-sided charge from the measured electric field flux, $\tilde{Q}_{tot}\propto \int_{\partial\Sigma_{+}}\star \tilde{F}_{A}+\int_{\partial\Sigma_{-}}\star \tilde{F}_{A}$, we find that the two contributions are equal rather than opposite; and hence, instead of the total 2-sided charge canceling (as in the black hole case) it is non-zero, and twice the charge of a one-sided black hole: $\tilde{Q}_{tot}\propto 2\int_{\partial\Sigma_{+}}\star \tilde{F}_{A}=2Q_{{\rm 1-sided}}$.  

In other words, since the black mirror's time orientation is flipped in exterior $I'$ (relative to the black hole), its charge is also flipped in exterior $I'$ (relative to the black hole).  This is consistent with the point, first emphasized by Stueckelberg \cite{Stueckelber1941, Stueckelberg:1941kpw, Stueckelberg:1941rg}, that if an electric charge is $-Q$ according to an observer with upward-pointing time orientation, it is $+Q$ according to an observer with downward-pointing time orientation ({\it i.e.} an anti-particle is a particle moving backward in time).

A closely analogous argument holds for the {\it mass}: if we use Gauss's law to determine the mirror's total (2-sided) Komar mass $\tilde{M}_{tot}$ by starting with the Killing one-form $k=-(1-\frac{2m}{r})dt$ (the one-form dual to the timelike Killing vector $\partial_{t}$), computing its ``field strength" $\tilde{F}_{k}=dk$, and finding $\tilde{M}_{tot}\propto \int_{\partial\Sigma_{+}}\star \tilde{F}_{k}+\int_{\partial\Sigma_{-}}\star \tilde{F}_{k}$, the contributions from both sides of the mirror are now {\it equal} (rather than opposite, as they were for the black hole).  In other words: in contrast to the Kruskal black hole, {\it the black mirror's Komar mass is {\bf positive} in both exteriors}, as expected from the positive-mass theorem \cite{schoen1979positive,schoen1981positive} -- a much healthier state of affairs!  

Now let us return to the surprising situation that, on the one hand, the black mirror's total 2-sided electric charge $\tilde{Q}_{tot}$ (or total 2-sided mass $\tilde{M}_{tot}$), as determined by the total outward-pointing electric (or gravitational) flux, is non-zero; yet, on the other hand (and contrary to Gauss's Law), there is no ``enclosed" charge (or mass) confined to the black mirror ({\it i.e.}\ no delta-function stress energy tensor $T_{\mu\nu}$ or current density $j^{\mu}$ living on the mirror at $\sigma=0$), since the black mirror remains a smooth solution of the vacuum Einstein-Maxwell equations across $\sigma=0$ (and, all components of the Riemann curvature $R^{\alpha}_{\;\;\beta\gamma\delta}$, Ricci tensor $R_{\mu\nu}$, and Einstein tensor $G_{\mu\nu}$ remain smooth and finite across $\sigma=0$).
Thus, the black mirror is a new incarnation of Misner and Wheeler's idea of ``mass without mass" and ``charge without charge" and a new illustration of the previously-known fact that Stokes' Theorem can be violated in certain topologically non-trivial circumstances \cite{Misner:1957mt, Sorkin:1977wq, Giulini:2009ts}.  In algebraic geometry, topologically nontrivial algebraic varieties like our black mirror are studied using standard ``blow-up" techniques \cite{griffiths1978principles}, and in follow-up work we are studying the black mirror from this standpoint to elucidate  this interesting behavior.

Finally, consider a particle with electric charge $q$ following the trajectory $x^{\mu}(\tau)$ in the region $\sigma>0$ of the charged Kerr black mirror, and the image of this trajectory under the $CPT$ map $\varphi_{CPT}$, in the region $\sigma<0$. Given that the charge of the black mirror -- as measured by an observer in $\sigma>0$ using Gauss' law -- is the same as the charge measured by an observer in $\sigma<0$, the invariance of the particle equations of motion under $CPT$ means that the charge of the particle as measured by an observer in $\sigma>0$ must also be the same as the charge of the CPT-image particle measured by an observer in $\sigma<0$, as shown in Figs.~\ref{collapsed-mirror}, \ref{chargeflow}.  Similarly, the unusual topology in the stationary case (discussed in the preceding paragraphs) suggests that, although the two infalling ``mirror particles" have he same charge, they nevertheless annihilate when they meet at the horizon.  (This is also the picture obtained from smoothly extending the solution across the horizon, in Appendix \ref{geodesics}.) In follow-up work (mentioned above), we apply blow-up techniques to the black mirror to help clarify this fascinating situation.

\section{Geodesics}
\label{geodesics}

The geodesics of infalling particles in the black mirror spacetime are illustrated in Fig.~\ref{collapsed-mirror}.  In this appendix we show that, working in the coordinates $(v_+,\sigma,\theta,\varphi)$ that smoothly cross the black horizon, massive particle geodesics may be smoothly extended across the horizon to give mirror-symmetric trajectories like those illustrated in Fig.~\ref{collapsed-mirror}.

For the sake of brevity, throughout this appendix we drop the subscript on $v_+$ and adopt units where the Schwarzchild radius $2m$ is unity. The line element is then
\begin{equation}
    \label{SchwarzschildEF}
    ds^{2}=-f(r)dv^2 +2 dv dr + r^2 d\Omega^2
\end{equation}
with $f=1-1/r$.
Since $k=\partial_{v}$ is a Killing vector, we have the conserved energy 
\begin{equation}
  \label{E}
  E=-k_{\mu}\dot{x}^{\mu}=f(r)\dot{v}-\dot{r},
\end{equation}
with $E$ positive ($E<1$ and $E>1$ for a bound and an unbound orbit, respectively).  If, for simplicity, we restrict attention to radial geodesics and ignore angular motion, the mass-shell constraint $\dot{x}_{\mu}\dot{x}^{\mu}=-1$ for a massive particle becomes
\begin{equation}
  \label{rdot}
  \dot{r}^{2}=E^{2}-f(r).
\end{equation}
Combining Eqs.~(\ref{E}, \ref{rdot}) we find, for an infalling particle,
\begin{equation}
  \frac{dv}{dr}=\frac{1-E/\sqrt{E^{2}-f(r)}}{f(r)},
\end{equation}
which may be integrated to yield
\begin{eqnarray}
    \label{RadialGeodesic}
    v&=&v_{0}+r-{r^{1\over 2} E \sqrt{1+r(E^2-1)}\over E^2-1}\cr
    &+&{E (3-2 E^2)\, {\rm sinh}^{-1}(r^{1\over 2} \sqrt{E^2-1} )\over   (E^2-1)^{3\over 2}}\cr
    &+&2 \ln ( r^{1\over 2} E +\sqrt{1+r(E^2-1)}).
\end{eqnarray}
where $v_{0}$ is an arbitrary constant. Now, if we re-express $r$ in terms $\sigma$ (in the units used in this appendix, $r=1+{1\over 4} \sigma^2$), we find that $v(\sigma)$ is a smooth, even function of $\sigma$, which ``turns around" at horizon crossing.\footnote{The explicit expression may prove useful in computing path integral propagators for quantum particles in a black mirror spacetime.} This conclusion is unchanged when angular motion is included, for any infalling geodesic intersecting the black horizon.

Note that, for a point $P$ on the black horizon, the only points in its causal future are other (later) points on the horizon (and in particular, those on the null generator passing through $P$).  So the correct physical interpretation of these extended (two-sided) geodesics is not that they describe particles passing from one exterior into the other; but rather, that (as shown in Figs.~\ref{collapsed-mirror}c and \ref{collapsed-mirror}f) they describe a particle and its CPT mirror image meeting (and annihilating) at the black horizon.  This, in turn, is reflected in the fact that, although the full (two-sided) geodesic is smoothly parameterized by $\sigma$, it is {\it not} smoothly parameterized by the proper time $\tau$ (since, as indicated by the arrows in Fig.~\ref{collapsed-mirror}), the proper time flows {\it toward} the black mirror on both sides.

\section{Saddle points of the action}
\label{saddle-point-app}
In this appendix, we show that the Schwarzchild black mirror presented in Section \ref{Schwarzschild} is a regular saddle of the Einstein-Hilbert action
\begin{equation}
  \label{EHaction}
{\cal S}=\int_{\cal M} dV {R\over 16 \pi} + {\rm boundary}\, {\rm terms}.
\end{equation}
The boundary terms are needed to ensure that the variation of the action yields the Einstein equations when the metric is held fixed on the boundary, with no constraint on its normal derivative there. The Einstein-Hilbert action does not have this property because it involves second derivatives of the metric. However, it can be reduced to an action involving only first derivatives by performing integrations by parts and discarding the boundary terms. The corrected action is then the appropriate one for a quantum path integral between an initial and a final three-geometry. 

A general principle, discussed by Coleman~\cite{Coleman:1985rnk}, states that if an action respects a symmetry $G$ then saddles of the action restricted to configurations which respect $G$ are also saddles of the full action. Hence, saddles of the gravitational action restricted to spherically symmetric spacetimes are also saddles of the full gravitational action. Consider the most general spherically symmetric line element,
\begin{equation}
  \label{spherical}
ds^2=-a(t,x)^2 dt^{2} + 2 S(t,x) dt dx +L(t,x)^2 dx^2 +r(x)^2 d\Omega^{2}.
\end{equation}
Here, $t$ and $x$ are $O(3)$-invariant coordinates. We do not assume the metric is static and include an arbitrary ``shift" $S(t,x)$ as well as a ``lapse" function $L(t,x)$. One can use the residual coordinate freedom to set $S(t,x)$ equal to zero and to choose $r$ as the radial coordinate. However, $r$ is not a good coordinate for our analytically extended spacetime and it is better to regard it as a function of a coordinate $x$ which is. 

In order to recover the full set of Einstein equations governing such metrics, one must vary the gravitational action before further restricting the coordinates. As we shall see, varying the action with respect to the shift $S$ yields a constraint that the lapse $L$ depends only on $x$. (This is the crux of Birkhoff's theorem, that every spherically symmetric solution of the Einstein equations is static.) Similarly, varying the action with respect to $L$ yields a constraint on $a$ and $r$ and their first $x$ derivatives. After obtaining these constraints, we can redefine the time coordinate to eliminate $S$ and reparameterize $x$ to set $L$ to unity. Finally, by varying the action with respect to $a$ and $r$ we obtain the second order field equations. These imply that $a(t,x)$ has a trivial time dependence which can be removed by reparameterizing $t$. 

To calculate the action (\ref{EHaction}) restricted to spherical symmetric spacetimes, we insert the ansatz (\ref{spherical}) into the Einstein-Hilbert term and integrate by parts to eliminate second derivatives. We need only retain terms of first order in $S(t,x)$ since it is set to zero after calculating the first order variations. After several integrations by parts, the action reduces (up to terms of higher order in $S$) to 
\begin{equation}
  \label{correctaction}
{\cal S}= 
\frac{1}{4}\int dt dx \left(a L +{r'(a r'+2ra')\over L}+2 S{rr'\dot{L}\over a L^2} \right),
\end{equation}
where dot denotes $\partial_t$ and prime $\partial_x$.  
Note that, we can derive this expression by simply extending the integrand (including $\sqrt{-g}$) smoothly from $x>0$ to $x<0$; or, equivalently, we can use the positive square root of $\sqrt{-g}$ if we also {\it reverse} the orientation of the $(\theta,\varphi)$ 2-sphere integration as we pass from $x>0$ to $x<0$; see Appendix \ref{CPT_appendix}. Setting the variation with respect to $S$ zero yields the constraint that $\dot{L}=0$.  (As will be seen later, in the solution $r'/a$ is finite and non-vanishing everywhere.) Varying the action with respect to $L$ and then setting it to unity yields the constraint
\begin{equation}
  \label{con1}
2 r r' a'=a(1-{r'}^2).
\end{equation}
Varying the action with respect to $a$ and $r$ then yields the two second order field equations
\begin{subequations}
  \begin{eqnarray}
   \label{eom1}
    r''&=& (1-{r'}^2)/(2 r),\\
     \label{eom2}
  a'' &=&-(ar')'/r.
  \end{eqnarray}
\end{subequations}
Eq.~(\ref{eom1}) is solved by noticing that $r({r'}^2-1)$ is constant, so that  ${r'}^2=1-2m/r$ with $m$ a constant. This implies $r''=m/r^2$. Now (\ref{con1}) yields $a'/a=r''/r'$ which is integrated to obtain $a=A(t)r'$ with $A(t)$ an arbitrary function. Assuming it is nonzero, it can be set to unity by redefining the time coordinate via $\tilde{t}=\int^t dt A(t)$. 

We now impose CPT symmetric boundary conditions. As we shall explain momentarily, $CPT$ will imply that $a(x)$ is an odd function of $x$.
To find the solutions explicitly, we set
\begin{equation}
  \label{chidef}
x= m\left(2 \chi+\sinh (2 \chi)\right ).
\end{equation}
The solution is then $r=2m\,{\rm cosh}^{2}\chi$, $a=\tanh \chi$ so that, in the $(t,\chi)$ coordinate system, the line element is
\begin{equation}
  \label{nicemetric}
ds^2=-(\tanh \chi)^2 dt^2 +(\cosh \chi)^4 \left((4m)^2 d\chi^2 +(2 m)^2d \Omega^{2}\right),
\end{equation}
with $-\infty <t,\chi <\infty$. The time-like one-form $e^0$ of the tetrad is odd in $\chi$ and has a simple zero at $\chi=0$, whereas the space-like one-forms $e^i$ are even and nonzero throughout. The spatial metric is conformal to a cylinder $\mathbb{R}\times S^2$, of radius $2m$. In the near-horizon limit ($|\chi|\ll 1$) the coordinate $\chi \approx \sigma/(4 m)$ where $\sigma$ is the proper radial coordinate used in Eqs.~(\ref{r(sigma)}--\ref{BlackMirrorSchwarzschild}); whereas far from the horizon ($\chi\rightarrow \pm \infty$) the line element (\ref{nicemetric}) becomes $\approx -dt^2 +m^2 e^{\pm 4 \chi}\left(d\chi^2 +{1\over 4} d \Omega^{2}\right)$
which, with $r={m\over 2} e^{\pm 2 \chi}$, is just Minkowski spacetime. 

Note that, the total action (\ref{correctaction}) -- which is the sum of the Einstein-Hilbert bulk action and the Gibbons-Hawking-York boundary terms -- vanishes when we integrate $dx$ over a symmetric range from $-X$ to $+X$, because the integrand is an odd function of $x$.

The method used here can be straightforwardly generalized to spherically symmetric electro-vacuum solutions that are asymptotically flat, de Sitter or anti-de Sitter. This time, after using various equations of motion obtained from the action, the line element takes the form
\begin{equation}
    ds^2 = -a^2(x)\,dt^2 + dx^2 +r^2(x)d\Omega^2,
\end{equation}
with $a=r'$ and 
\begin{eqnarray}
    {r'}^2= 1-\frac{2m}{r}+\frac{q_e^2}{r^2} - \Lambda r^2.
\end{eqnarray}
The function $a=a(r)$ has at most three roots: $r_-$ which corresponds to the Cauchy horizon, $r_+$ which corresponds to the event horizon and $r_c$ which corresponds to the cosmological horizon, with $r_c = +\infty$ if $\Lambda\leq 0$. Since $a^2\geq 0$ it follows that the range of $r$ lies in the interval $[r_+,r_c]$. Moreover, CPT symmetry implies that $r$ obeys $r'(0)=0$, and therefore $r(0)$ is a root of $a=a(r)$. If $r(0) = r_c$ then the function $r=r(x)$ is a constant. Thus, the only physical initial conditions for $r$ are $r(0)=r_+$ and $r'(0)=0$. Extending this solution to the horizon by changing coordinates from Schwarzschild time $t$ to Eddington-Finkelstein time $v$ ($v_{+}$ or $v_{-}$) again brings us back to the black mirror discussed in the main text.  

\end{appendix}

\end{document}